\documentclass[11pt]{article}

\usepackage{amsfonts} 
\usepackage[utf8]{inputenc}
\usepackage[super]{cite}

\title{The Scalar Mode of Gravity}

\author{
Metin Arık, Tarik Tok \\
  Department of Physics\\
 Bogazici University\\
   Bebek, Istanbul, Turkey\\
  \texttt{metin.arik@boun.edu.tr}\\
  \texttt{tarik.tok@boun.edu.tr} }

\begin{document}

\maketitle
\begin{abstract}
We consider the scalar mode of gravity as expressed  by a conformal factor of the metric and present a model motivated by the Jordan-Brans-Dicke action. The reduced action provides a Lagrangian density in Minkowski space which exhibits a massive particle and an expanding space-time through a mechanism which is similar to the Higgs mechanism. 
\end{abstract}

 Einstein's theory of gravity is perhaps the most extraordinary achievement of modern physics. General relativity\cite{einstein1,einstein2,einstein3} not only explained the observed  perihelion advance of Mercury  and predicted  the correct  bending  of a light ray passing through a nearby point on the surface of the Sun but also foreshadowed the expanding universe and gravitational waves. From a modern point of view Einstein's equation in vacuum  can be derived considering the Pauli-Fierz field theory\cite{fierz}  for a massless spin-2 graviton, coupling the canonical energy momentum tensor as a source of the graviton and iterating to all orders which  is a lengthy derivation\cite{gupta,feynmann} as compared to the brilliant Einstein argument of covariance and conservation. Although gravitational waves have been detected by an experiment\cite{ligo} which is the most sensitive to the spin-2 nature of the graviton, it cannot yet rule out spin-0 gravitational waves.

Although vacuum  Einstein equations unquestionably describe a spin-2 massless particle as the carrier of the gravitational force, the FLRW model of cosmology\cite{friedman1,friedman2,lematre1,lematre2,robertson,walker}  includes only a scalar field which is the scale size of the universe and is only time dependent due to the idealization of isotropy  and homogeneity. This paper concentrates on this scalar mode and investigates how much can be explained by neglecting the tensor mode of general relativity. Historically, first Einstein and then Nordstrom\cite{Nordstorm,Nordstorm1,Nordstorm2} initially considered such theories, but they were later abandoned due to the successes of Einstein's General Relativity. Scalar fields in physics have gained renewed importance due to the triupmh of the standard model\cite{weinberg,abdussalam,glashow} of particle physics which says that the masses of all the massive elementary particles (the leptons, the quarks and the weak bosons) are given by a dimensionless coupling times the vacuum expectation value of the Higgs field. Although this value is time independent  in the standard model, it is natural to consider that it may well depend on the cosmological time scale t, so that the trajectory of a particle in cosmic time scales is given by the geodesic action  

\begin{equation} \label{eq1}
S=\int{}{}m_0\frac{\Phi (t)}{\Phi_0} \bigg(dt^2-a^2_0 \frac{d\vec{r}^2}{\big(1+\frac{k\vec{r}^2}{4}\big)^2} \bigg)^\frac{1}{2}
\end{equation}
where $m_0$ is the present mass of the particle, $\Phi(t)$ is the cosmological vacuum expectation value of the Higgs field and $\Phi_0$ is its present value with $a_0$ being the present scale size of the universe. One immediate question which arises is what principle determines $\Phi(t)$. The most natural choice for determining this is the Einstein–Hilbert action\cite{hilbert} with the reduced Lagrangian

\begin{equation}
L=\frac{1}{2} \dot{\Phi}^2-\frac{1}{2} \frac{k}{a^2_0}\Phi^2.
\end{equation}
Without any source term for cosmology, this describes a scalar particle whose Compton wavelength is same as the size of the universe, $a_0$, for $k=+1$.

More generally the field $\Phi$ can also depend on the space variables and for a universe with $k=0$, one can write a more general Lagrangian density which is motivated by scale invariant Jordan-Brans-Dicke theory\cite{jordan,jordan2,brans-dicke}. This Lagrangian density is given by

\begin{equation}\label{eq5}
\mathcal{L}=\bigg(\frac{1}{2} \zeta^2 \chi^2 R+\frac{1}{2} g^{\mu\nu} \partial_\mu\chi \partial_\nu\chi-\frac{1}{4} \lambda \chi^4 \bigg) \sqrt{- \det g}
\end{equation}
where 
$g_{\mu\nu}=\Phi^2(x) \eta_{\mu\nu}$, $\chi$ is Jordan scalar field and $\zeta^2$ is proportional to the inverse Brans-Dicke parameter \cite{kurkov,kondo,kirsh,bauerdademir,Horne,germani}. Quartic interaction term for the Jordan field has been added to the model without violating scale invariance. The Lagrangian density which is thus motivated by the conformally flat metric in (\ref{eq1}) and the above equation is given by

\begin{equation} \label{eq2}
\mathcal{L}= \frac{1}{2} \zeta^2 \chi^2 \partial_\mu\Phi\partial^\mu\Phi+\frac{1}{2}\Phi^2\partial_\mu\chi\partial^\mu\chi-\frac{\lambda}{4} \chi^4\Phi^4
\end{equation}
where, now the indices are lowered and raised by the Minkowski metric. Note that although (\ref{eq5}) defines a field theory in curved spacetime with metric $g_{\mu\nu}$, (\ref{eq2}) defines a field theory in Minkowski space.

We take the scalar fields to have scalar dimensions of $(mass)^\frac{1}{2}$ so that the parameter $\zeta$ and $\lambda$ are dimensionless. The Minkowski metric is defined as

$$\eta_{\mu\nu}=(+1,-1,-1,-1)$$
and we use units where $\hbar=c=1$.
\indent
 
 The Lagrangian density(\ref{eq2})  has SO(1,1) symmetry acting on $\Phi$ and $\chi$ as $\Phi \rightarrow a\Phi$ , $\chi \rightarrow \frac{1}{a} \chi$, where a is constant. Due to this invariance the Lagrangian simplifies
 when new field variables $\alpha$ and $\gamma$ are defined by
 
 \begin{equation}  \Phi=\alpha^{\frac{1}{2}} e^{-\gamma}\end{equation} 
 \begin{equation} \chi=\alpha^{\frac{1}{2}} e^{\gamma}\end{equation} 
 so that $\alpha$ has the dimension of mass and $\gamma$ is dimensionless. We will show that this Lagrangian leads to spontaneous symmetry breaking\cite{srednicki} even though Lagrangian density does not contain any dimensional parameters.
 
Under the assumption that the fields depend only on time, the Lagrangian density (\ref{eq2}) in terms of $\alpha$ and $\gamma$ becomes

\begin{equation}
\mathcal{L}=\frac{1}{8}(1+\zeta^2)\dot{\alpha}^2+\frac{1}{2}(1+\zeta^2)\alpha^2\dot{\gamma}^2+\frac{1}{2}(1-\zeta^2)\alpha \dot{\alpha}\dot{\gamma}- \frac{\lambda}{4}\alpha^4.
\end{equation}

 Since $\mathcal{L}$ does not contain the field $\gamma$, but only its derivatives, it leads to the conserved momentum 

\begin{equation}\label{eq3}
\frac{\partial\mathcal{L}}{\partial\dot{\gamma}}=\frac{1}{2}(1-\zeta^2)\alpha \dot{\alpha}+(1+\zeta^2)\alpha^2 \dot{\gamma}=\mu^3
\end{equation}
where the conserved momentum which has dimension of $(mass)^3$ has been equated to $\mu^3$.

Since the observed space-time is expanding on the cosmological scale, we assume that the cosmological expectation values of $\alpha$ and $\gamma$ may also depend on the time variable t.

The Hamiltonian density becomes

\begin{equation}\label{4}
\mathcal{H}=\frac{1}{2} \frac{\zeta^2}{(1+\zeta^2)} \dot{\alpha}^2+\frac{\mu^6}{2(1+\zeta^2)} \frac{1}{\alpha^2}+\frac{\lambda}{4}\alpha^4.
 \end{equation}
Thus, the effective potential for our dynamical system is given by 

\begin{equation}
V_{eff}=\frac{\mu^6}{2(1+\zeta^2)} \frac{1}{\alpha^2}+\frac{\lambda}{4}\alpha^4.
\end{equation}
The minimum of effective potential is given by 

\begin{equation}
\alpha_0=\frac{\mu}{\big(\lambda(1+\zeta^2)\big)^\frac{1}{6}}
\end{equation}
and, for $\alpha=\alpha_0$, $\gamma$ is given from (\ref{eq3})

\begin{equation}
\gamma_0(t)=C+Dt
\end{equation}
where C is an integration constant and 

\begin{equation}
D=\frac{\mu^3}{(1+\zeta^2)\alpha^2_0}.
\end{equation}
Now, we expand $\alpha$ and $\gamma$ about $\alpha_0$ and $\gamma_0(t)$ by

\begin{equation}\label{eq14}
\alpha=\alpha_0(1+\epsilon A e^{imt}+\epsilon A^* e^{-imt})+\mathcal{O}(\epsilon^2)
 \end{equation}

\begin{equation}\label{eq15}
\gamma=\gamma_0(t) +\epsilon E e^{imt}+\epsilon E^*  e^{-imt}+\mathcal{O}(\epsilon^2)
\end{equation}
where, $\epsilon$ is a small dimensionless parameter.

After performing a perturbative calculation neglecting higher powers of $\epsilon$, we find that for consistency the mass $m$ and frequency  in  (\ref{eq14}) and (\ref{eq15}) should be given by

\begin{equation}\label{eq16}
m=\sqrt{6}\frac{(1+\zeta^2)^\frac{1}{3}}{\zeta}  \lambda^\frac{1}{3} \mu.
\end{equation}

Furthermore, we find that $A$ and $E$ are not independent but related by 

\begin{equation}
E=\bigg(i\frac{2}{\sqrt{6}}\frac{\zeta}{(1+\zeta^2)}-\frac{1}{2} \frac{(1-\zeta^2)}{(1+\zeta^2)}+ \mathcal{O}(\epsilon)\bigg)A
\end{equation}
in lowest order.

The part, $C+Dt$, in (\ref{eq15}) which in quantum field theory would describe a massless particle, in classical physics corresponds to an expanding universe.

We note that the system can be quantized by imposing the commutation relation

\begin{equation}\label{eqqu1}
\left[ A,A^*\right]=1  
\end{equation}
so that $A$ and $A^*$ become the creation and annihilation operators of the $\alpha$-particle. On the other hand, $E$ and $E^*$ which are not independent of $A$ and $A^*$ can be interpreted as quantum corrections to $\gamma_0(t)$ which governs the expansion of the universe. This expansion is obtained by taking the average value of $\gamma(t)$ 

\begin{equation}\label{eqnn19}
\langle \gamma \rangle =C+Dt
\end{equation}
which gives

\begin{equation}\label{eq20}
\langle \Phi \rangle =\alpha_0^\frac{1}{2} e^{-C-Dt}= \Phi_0 e^{-Dt}
\end{equation}

\begin{equation}
\langle \chi \rangle =\alpha_0^\frac{1}{2} e^{C+Dt}= \chi_0 e^{Dt}.
\end{equation}
Equation (\ref{eq20}) together with the metric(\ref{eq1}) defines (for $D<0$) a space-time which is exponentially expanding in the conformal time variable t. When transformed to cosmological time, this gives a linearly expanding space. Thus we have been able to obtain a massive field with mass m and an effectively linearly expanding space(\ref{eqnn19}) by using the Lagrangian density(\ref{eq2}) in Minkowski space. Although building a realistic unified theory of cosmology  and the standard model along the lines of this paper seems questionable, it is never the less a useful exercise to contemplate that the expanding universe and the Higgs mechanism  can be considered as two faces of the same coin presented in this paper. Metin Arık would like to thank the Turkish Academy of Sciences for support and acknowledge numerous discussions with Tolga Yarman.

\bibliography{references}

\end{document}